# Adaptive Sparse Channel Estimation for Time-Variant MIMO-OFDM Systems

Guan Gui, Wei Peng, and Fumiyuki Adachi
Department of Communication Engineering
Graduate School of Engineering,
Tohoku University
Sendai, Japan
{gui,peng}@mobile.ecei.tohoku.ac.jp, adachi@ecei.tohoku.ac.jp

*Abstract*—**Accurate channel state information (CSI) is required for coherent detection in time-variant multiple-input multiple-output (MIMO) communication systems using orthogonal frequency division multiplexing (OFDM) modulation. One of low-complexity and stable adaptive channel estimation (ACE) approaches is the normalized least mean square (NLMS)-based ACE. However, it cannot exploit the inherent sparsity of MIMO channel which is characterized by a few dominant channel taps. In this paper, we propose two adaptive sparse channel estimation (ASCE) methods to take advantage of such sparse structure information for time-variant MIMO-OFDM systems. Unlike traditional NLMS-based method, two proposed methods are implemented by introducing sparse penalties to the cost function of NLMS algorithm. Computer simulations confirm obvious performance advantages of the proposed ASCEs over the traditional ACE.**

## I. INTRODUCTION

The use of multiple-input multiple-output (MIMO) transmission (as shown in Fig. 1), and orthogonal frequency division multiplexing (OFDM) makes high data communications possible with low transmit power in a frequency-selective fading channel [1–3]. In the high mobility environment, the MIMO channel is subjected to a as time-variant fading (i.e., double-selective fading). The accurate estimation of channel impulse response (CIR) is a crucial and challenging issue in coherent modulation and its accuracy has a significant impact on the overall performance of communication system.

During last decades, many channel estimation methods proposed for MIMO-OFDM systems [4–12]. However, all of the proposed methods can be categorized into two types. The first type is that linear channel estimation methods, e.g., least squares (LS) algorithm, based on the assumption of dense CIRs. By applied these approaches, the performance of linear methods depend only on size of MIMO channel. Note that narrowband MIMO channel may be modeled as dense channel model because of its very short time delay spread; however,

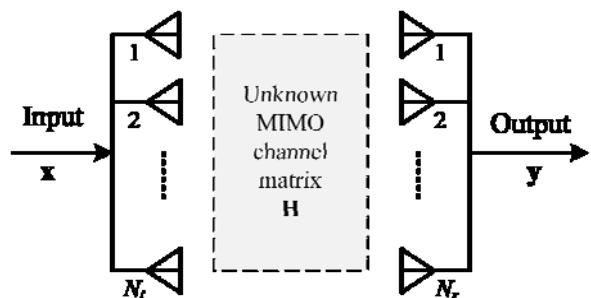
Fig.1. An example of MIMO system.

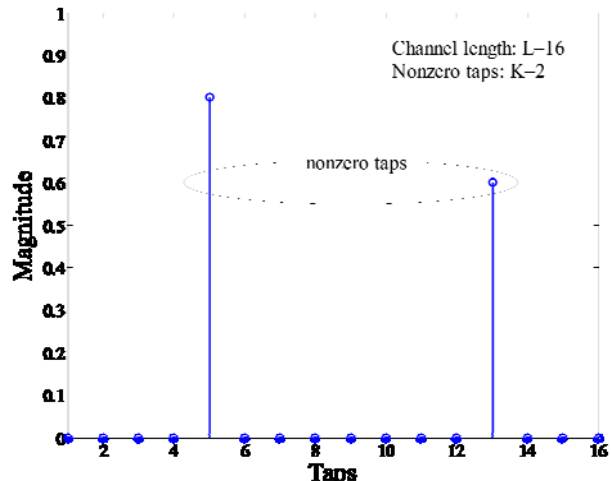
Fig. 2. A typical example of sparse multipath channel.

broadband MIMO channel is often modeled as sparse channel model [13–15]. A typical example of sparse channel is shown in Fig. 2. It is well known that linear channel estimation methods are relatively simple to implement due to its low computation complexity [4–9]. But, the main drawback of linear channel estimations is unable to exploit the inherent channel sparsity. Second type is the sparse channel estimation methods using compressive sensing (CS) [16], [17] based on the assumption of sparse CIRs. Optimal sparse channel estimation often requires that its training signal satisfies restrictive isometry property (RIP) [18]. However, designing the RIP-satisfied training signal is a Non-Polynomial (NP) hard problem [19]. Although some CS algorithms are stable solution, they incur extra high computational burden, especially in time-variant MIMO-OFDM systems. For example, one of typical sparse channel estimations methods, using Dantzig selector (DS) algorithm, was proposed for double-selective fading MIMO systems in [11]. However, DS algorithm needs to be solved by linear programming and hence it incurs high computational complexity. To reduce the complexity, sparse channel estimation methods using greedy iterative algorithms were also proposed in [10], [12]. However, their complexity depends on the number of nonzero taps of MIMO channel. Larger number of nonzero taps in MIMO channel requires higher complexity and vice versa. Hence, a new sparse channel estimation method needs to be developed for MIMO-OFDM channel.

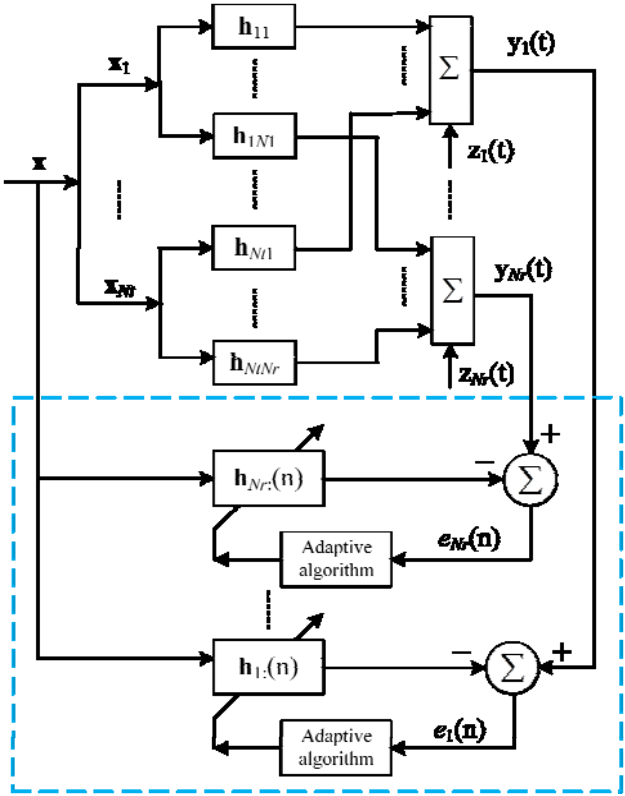

Fig.3. ASCE for MIMO-OFDM systems.

In this paper, we study least mean square (LMS) [20] based adaptive channel estimation (ACE) method for MIMO-OFDM systems. To exploit the channel sparsity, Chen et. al. proposed an effective sparse LMS algorithm using $L_1$-norm sparse penalty [21]. Based on the $L_1$-norm sparse LMS, Taheri et. al. proposed a $L_p$-norm LMS (LP-LMS) based adaptive sparse channel estimation (ASCE) method to further exploit the channel sparsity in signal-antenna systems [22]. To fully take advantage of channel sparsity and to improve stability of adaptive parse channel estimation (ASCE), we proposed a $L_0$-norm LMS (L0-LMS) based ASCE method and sparse normalized LMS (NLMS) methods, i.e., $L_p$-norm NLMS (LP-NLMS) and $L_0$-norm NLMS (L0-NLMS) algorithms, for single-antenna time-variant communication systems [23]. To the best of our knowledge, it is still no work to propose ASCE method in MIMO-OFDM systems. Hence, we extend our previous work in [23] to MIMO systems. Our proposed ASCE methods are implemented by LP-NLMS and L0-NLMS, respectively. First of all, as shown in Fig. 3, MIMO-OFDM system model is formulated so that each multiple-input single-output (MISO) channel vector can be estimated by ASCE methods. Later, computer simulation results are presented to confirm the effectiveness of our proposed methods.

The remainder of this paper is organized as follows. A MIMO-OFDM system model is described and problem formulation is given in Section II. In section III, the sparse LMS algorithm is introduced and ASCE in MIMO-OFDM systems is highlighted. In addition, performances of ASCE methods are compared analytically. Computer simulation results are given in Section IV in order to evaluate and compare performances of the ASCE methods. Finally, we conclude the paper in Section V.

## II. SYSTEM MODEL

Consider a time-variant MIMO-OFDM communication system as shown in Fig. 1. Frequency-domain signal vector $\bar{\mathbf{x}}_{n_t}(t) = [\bar{x}_{n_t}(t,0), ..., \bar{x}_{n_t}(t, C-1)]^T$, $n_t = 1,2, ..., N_t$ is fed to inverse discrete Fourier transform (IDFT) at the $n_t$-th antenna, where $C$ is the number of subcarriers. Assume that the transmit power is $\{\|\bar{\mathbf{x}}_{n_t}(t)\|\} = CE_0$. The resultant vector $\mathbf{x}_{n_t}(t) \triangleq \mathbf{F}^H \bar{\mathbf{x}}_{n_t}(t)$ is padded with cyclic prefix (CP) of length $L_{CP} \geq (K-1)$ to avoid inter-block interference (IBI), where $\mathbf{F}$ is a $C \times C$ DFT matrix with entries $[\mathbf{F}]_{cq} = 1/K\, e^{-j2\pi cq/C}$, $c, q = 0,1, ..., C-1$. After CP removal, the received signal vector at the $n_t$-th antenna for time $t$ is written as $y_{n_r}$. Then, the received signal $\mathbf{y}$ and input signal $\mathbf{x}$ are related by

$$\mathbf{y} = \mathbf{Hz} + \mathbf{z}, \qquad (1)$$

where the MIMO channel matrix $\mathbf{H}$ can be written as

$$\mathbf{H} = \begin{bmatrix} \mathbf{h}_{11}^T & \mathbf{h}_{12}^T & \cdots & \mathbf{h}_{1N_t}^T \\ \mathbf{h}_{21}^T & \mathbf{h}_{22}^T & \cdots & \mathbf{h}_{2N_t}^T \\ \vdots & \vdots & \ddots & \vdots \\ \mathbf{h}_{N_r1}^T & \mathbf{h}_{N_r1}^T & \cdots & \mathbf{h}_{N_rN_t}^T \end{bmatrix} = \begin{bmatrix} \mathbf{h}_{1:}^T \\ \mathbf{h}_{2:}^T \\ \vdots \\ \mathbf{h}_{N_r:}^T \end{bmatrix}, \qquad (2)$$

where $N$ is the size of the channel memory of each single channel between each antenna pair. Then, the received signal at $n_r$-th antenna can be written as

$$y_{n_r} = \sum_{n_t=1}^{N_t} \mathbf{h}_{n_r n_t}^T \mathbf{x}_{n_t} + z_{n_r} = \mathbf{h}_{n_r:}^T \mathbf{x} + z_{n_r}, \qquad (3)$$

where $\mathbf{h}_{n_r:}^T = [\mathbf{h}_{n_r1}^T, \mathbf{h}_{n_r2}^T, ..., \mathbf{h}_{n_rN_t}^T] \in \mathbb{C}^{1 \times N_t C}$, $n_r = 1,2, .., N_r$ is a column vector which is considered as an multiple input single output (MISO) channel vector and $\mathbf{h}_{n_r n_t}$ ($n_r = 1,2, ..., N_r$ and $n_t = 1,2, ..., N_t$) is assumed equal $L$-length sparse channel vector from $n_r$-th receiver antenna to $n_t$-th antenna. In addition, we assume that the each channel vector $\mathbf{h}_{n_r n_t}$ is only supported by $K$ dominant channel taps. A typical example of sparse multipath channel is depicted in Fig. 2. Hereby, at the $n_r$-th receive antenna, the corresponding signal estimation error $e_{n_r}$ at time $t$ can be defined as

$$e_{n_r}(n) = y_{n_r} - y_{n_r}(n) = y_{n_r} - \mathbf{h}_{n_r:}^T(n)\mathbf{x}(n), \quad (4)$$

for $n_r = 1,2,...,N_r$, where $\mathbf{h}_{n_r:}^T(n)$ denotes an $n_r$-th adaptive updating estimator of $\mathbf{h}_{n_r:}^T$ and $y_{n_r}(n)$ is the output signal from NLMS filter which can be seen in Fig.3. If we collect all of the error signals $e_{n_r}(n), n_r = 1,2,...,N_r$, then the Eq. (4) can be rewritten as matrix-vector form

$$\begin{aligned}\mathbf{e}(n) &= [e_1(n), e_2(n), ..., e_{N_r}(n)]^T \\ &= \mathbf{y} - \mathbf{y}(n) \\ &= \mathbf{y} - \mathbf{H}(n)\mathbf{x}(n),\end{aligned} \quad (5)$$

where $\mathbf{y} = [y_1,...,y_{N_r}]^T$ and $\mathbf{y}(n) = [y_1(n),...,y_{N_r}(n)]^T$ denote MIMO system ideal output vector and its estimate signal, respectively; $\mathbf{H}(n)$ is an $n$-th adaptive estimate channel matrix $\mathbf{H}$. According to Eq. (5), MIMO channel estimation problem equivalents to estimate different individual MISO channel $\mathbf{h}_{n_r:}$ using error signal $e_{n_r}(n)$ and input training signal $\mathbf{x}(n)$. In general, estimate the MISO channel vector $\mathbf{h}_{n_r:}$ using standard LMS algorithm, the corresponding cost function can be constructed as

$$L_{n_r}(n) = \tfrac{1}{2}e_{n_r}^2(n), \quad (6)$$

for $n_r = 1,2,...,N_r$. It is obvious that the update equation of LMS based adaptive channel estimation can be derived as

$$\begin{aligned}\mathbf{h}_{n_r:}(n+1) &= \mathbf{h}_{n_r:}(n) - \mu \tfrac{\partial L_{n_r}(n)}{\partial \mathbf{h}_{n_r:}(n)} \\ &= \mathbf{h}_{n_r:}(n) + \mu e_{n_r}(n)\mathbf{x}(n)\end{aligned} \quad (7)$$

for $n_r = 1,2,...,N_r$, where $\mu \in (0, \gamma_{\max}^{-1})$ is the step size of LMS gradient descend and $\gamma_{\max}$ is the maximum eigenvalue of the covariance matrix $\mathbf{R} = E\{\mathbf{x}(t)\mathbf{x}^T(t)\}$. Since the LMS based method is sensitive to random scaling of training signal. To improve the stability, normalized LMS (NLMS) is considered as standard method for MIMO ACE. Hence, its update equation is given by

$$\mathbf{h}_{n_r:}(n+1) = \mathbf{h}_{n_r:}(n) + \mu \frac{e_{n_r}(n)\mathbf{x}(n)}{\mathbf{x}^H(n)\mathbf{x}(n)}. \quad (8)$$

### III. PROPOSED SPARSE NLMS METHODS

Consider $L_p$-norm sparse penalty on NLMS cost function to produce sparse channel estimator since this penalty term forces the channel taps values of $\mathbf{h}_{n_r:}$ to approach zero. It is termed as LP-NLMS which was proposed for single-antenna systems in [22]. For $n_r$-th MISO channel vector, its cost function of the LP-NLMS is given by

$$L_{lp,n_r}(n) = \tfrac{1}{2}e_{n_r}^2(n) + \lambda_{lp,n_r}\|\mathbf{h}_{n_r:}\|_p, \quad (9)$$

for $n_r = 1,2,...,N_r$, where $\|\cdot\|_p$ is the $L_p$-norm operator and $\lambda_{lp,n_r}$ is a regularization parameter which tradeoffs the mean square error and sparse penalty. The update equation of LP-NLMS based adaptive sparse channel estimation can be derived as

$$\mathbf{h}_{n_r:}(n+1) = \mathbf{h}_{n_r:}(n) + \mu \frac{e_{n_r}(n)\mathbf{x}(n)}{\mathbf{x}^H(n)\mathbf{x}(n)} \\ -\rho_{lp,n_r}\frac{\|\mathbf{h}_{n_r:}(n)\|_p^{1-p}\mathrm{sgn}(\mathbf{h}_{n_r:}(n))}{\sigma + |\mathbf{h}_{n_r:}(n)|^{1-p}}, \quad (10)$$

where $\rho_{lp,n_r} = \mu\lambda_{lp,n_r}$ depends on gradient descend step-size $\mu$ and regularization parameter $\lambda_{lp,n_r}$.

Following to idea of the LP-NLMS algorithm on adaptive channel estimation, if $p = 0$, then the zero-attracting forces the channel taps values of $\mathbf{h}_{n_r:}$ to approach zero is L0-norm penalty. It is termed as L0-norm NLMS (L0-NLMS) [23] that the cost function is given by

$$L_{l0,n_r}(n) = \tfrac{1}{2}e_{n_r}^2(n) + \lambda_{l0,n_r}\|\mathbf{h}_{n_r:}(n)\|_0, \quad (11)$$

where $\|\cdot\|_0$ is the $L_0$-norm operator that counts the number of nonzero taps in $\mathbf{h}_{n_r:}(n)$ and $\lambda_{l0,n_r}$ is a regularization parameter to balance the estimation error and sparse penalty. Since solve the $L_0$-norm minimization is a NP-hard problem [19], we replace it with approximate continuous function

$$\|\mathbf{h}_{n_r:}\|_0 \approx \sum_{l=0}^{N_tN-1}\left(1 - e^{-\beta|h_{n_r:,l}|}\right). \quad (12)$$

According to the approximate function, L0-LMS cost function can be revised as

$$L_{l0,n_r}(n) = \tfrac{1}{2}e_{n_r}^2(n) + \lambda_{l0,n_r}\sum_{l=0}^{N_tN-1}\left(1 - e^{-\beta|h_{n_r:,l}|}\right), \quad (13)$$

Then, the update equation of L0-LMS based adaptive sparse channel estimation can be derived as

$$\mathbf{h}_{n_r:}(n+1) = \mathbf{h}_{n_r:}(n) + \mu e_{n_r}(n)\mathbf{x}(n) \\ -\rho_{l0,n_r}\beta\mathrm{sgn}\left(\mathbf{h}_{n_r:}(n)\right)e^{-\beta|\mathbf{h}_{n_r:}(n)|}, \quad (14)$$

where $\rho_{l0,n_r} = \mu\lambda_{l0,n_r}$. It is worth mention that the exponential function in (14) will cause high computational complexity. To reduce the computational complexity, the first order Taylor series expansion of exponential functions is taken into consideration as [24]

$$e^{-\beta|h|} \approx \begin{cases} 1 - \beta|h|, & \text{when } |h| \leq 1/\beta \\ 0, & \text{others.} \end{cases} \quad (15)$$

Then, the update equation of L0-NLMS based adaptive sparse channel estimation can be derived as

$$\mathbf{h}_{n_r:}(n+1) = \mathbf{h}_{n_r:}(n) + \mu \frac{e_{n_r}(n)\mathbf{x}(n)}{\mathbf{x}^H(n)\mathbf{x}(n)} - \rho_{l0,n_r}J(\mathbf{h}_{n_r:}(n)), \quad (16)$$

where $J(h)$ is defined as

$$J\left(\mathbf{h}_{n_r:}(n)\right) = \begin{cases} 2\beta^2 h - 2\beta\mathrm{sgn}(h), & \text{when } |h| \leq 1/\beta \\ 0, & \text{others.} \end{cases} \quad (17)$$

### IV. NUMERICAL SIMULATIONS

In this section, the proposed ASCE estimators using 1000 independent Monte-Carlo runs for averaging. The length of channel vector $\mathbf{h}_{n_tn_r}$ between each pair $(n_t, n_r)$ is set as $N = 16$ and its number of dominant taps is set as $K = 1$ and 4, respectively. Values of dominant channel taps follow Gaussian distribution and their positions are randomly allocated within the length of $\mathbf{h}_{n_tn_r}$ which is subjected to $E\{\|\mathbf{h}_{n_tn_r}\|_2^2 = 1\}$. The received signal-to-noise ratio (SNR) is defined as $10\log(E_0/\sigma_n^2)$, where $E_0 = 1$ is transmitted power at each antenna. Here, we set the SNR as 5dB, 10dB and 15dB in computer simulation. All of the step sizes and regularization parameters are listed in Tab. I. The estimation performance is

evaluated by average mean square error (MSE) which is defined as

$$Avergae\ MSE\{\mathbf{H}(n)\} = E\{\|\mathbf{H} - \mathbf{H}(n)\|_2^2\}, \quad (18)$$

where $E\{\cdot\}$ denotes expectation operator, $\mathbf{H}$ and $\mathbf{H}(n)$ are the actual MIMO channel vector and its $n$-th adaptive channel estimator, respectively.

TABLE I.  SIMULATION PARAMETERS.

| Parameters | $\mu$ | $\lambda_{lp}$ | $\lambda_{l0}$ |
|---|---|---|---|
| Values | 0.5 and 1 | $(1e-4)\sigma_n^2$ | $(1e-3)\sigma_n^2$ |

In the first example, the proposed methods are evaluated in Figs. 4-6 in different signal region. The three figures show that LP-NLMS based ASCE methods can achieve better estimation performance than standard NLMS based ACE. Since L0-NLMS based ASCE methods taken more fully sparsity advantage of MIMO channel, much better estimation performance than NLMS is achieved. In addition, the three figures also indicate that ACE using NLMS has no obvious relationship with number of nonzero channel taps; but proposed ASCE methods depend on the number of nonzero taps. The proposed ASCE methods can achieve better estimation performance for sparser channel, and vice versa. Let us take the Fig. 5 example. Two performance curves of ACE method using NLMS are almost same at different number of nonzero channel taps, i.e., $K = 1$ and 4. Unlike this phenomenon, ASCE methods using both LP-NLMS and L0-NLMS algorithms achieved better estimation performance on the channel ($K = 1$) which is sparser than the channel ($K = 4$).

In the second experiment, the proposed methods are evaluated at different number of transmit/receive antennas, i.e., $(N_t, N_r) = (2,4)$ as shown in Figs. 7-8. The figures are shown that the performance advantage of proposed ASCE using methods LP-NLMS and L0-NLMS algorithms over than SCE method using NLMS algorithm.

In the third experiment, the proposed methods are evaluated at different step-sizes, i.e., $\mu = 0.5, 1$ and $1.5$, as shown in Fig. 9-10. Without loss of generality, the simulation environment is considered in SNR = 10dB. The number of transmit/receive antennas are set as $(N_t, N_r) = (2,2)$ in Fig. 9 and $(N_t, N_r) = (2,4)$ in Fig. 10, respectively. The two figures show that the proposed ASCE methods can achieve better performance than standard ACE method. Here, note that each proposed ASCE method using smaller gradient descend step-size can achieve better estimation that one using bigger step-size, at the cost of few higher computational complexity and vice versa. For a practical time-variant MIMO-OFDM system, both performance and simplicity of the ASCE method are required. In different SNR region, hence, different step-size could be applied to trade off the performance and computational complexity of the proposed ASCE methods.

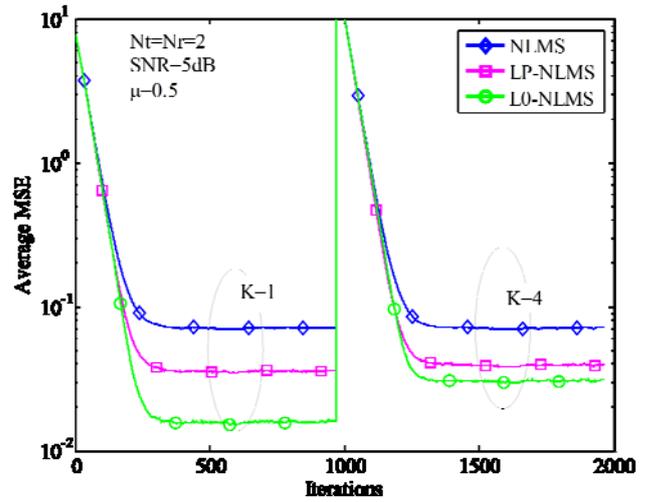
Fig. 4. Performance comparison at SNR = 5dB and $\mu = 0.5$.

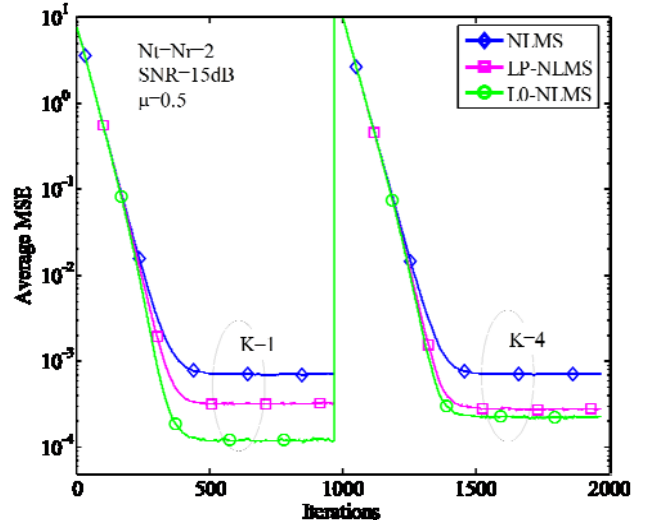
Fig. 5. Performance comparison at SNR = 10dB and $\mu = 0.5$.

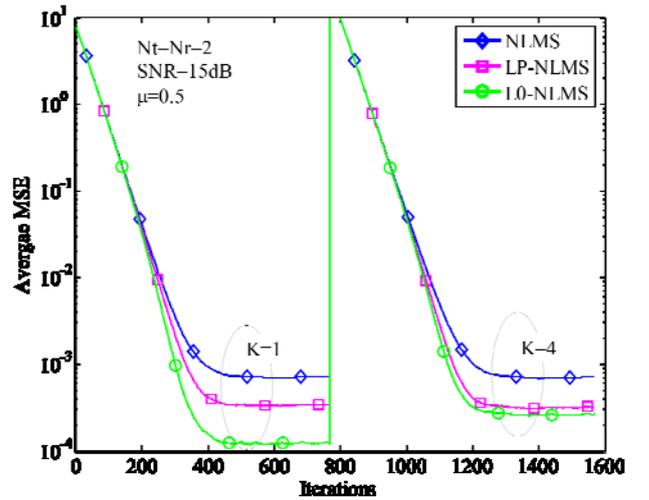
Fig. 6. Performance comparison at SNR = 15dB and $\mu = 0.5$.

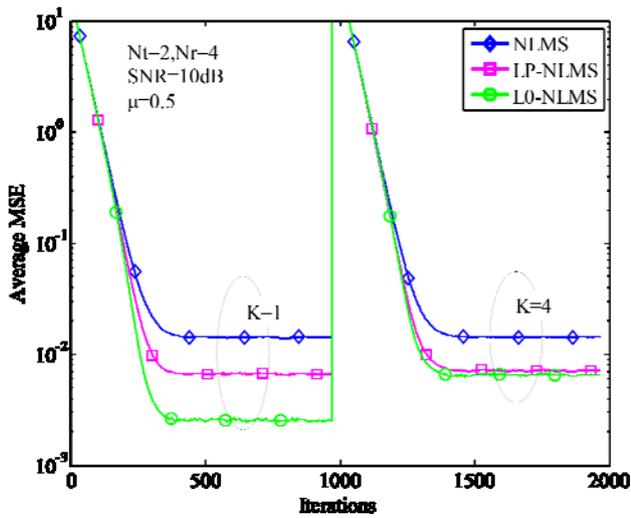

Fig. 7. Performance comparison at SNR = 10dB and $\mu = 0.5$.

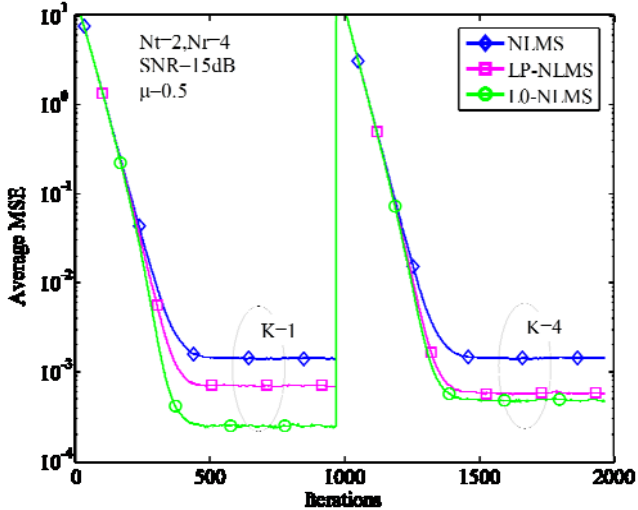

Fig. 8. Performance comparison at SNR = 15dB and $\mu = 0.5$.

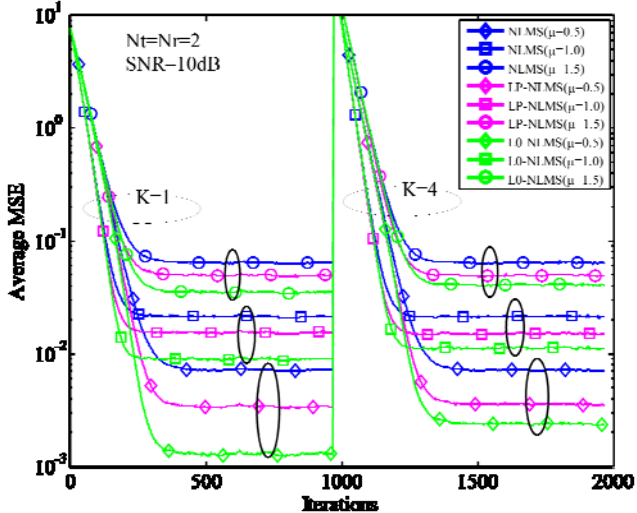

Fig. 9. Performance comparison verses step-size of gradient descend.

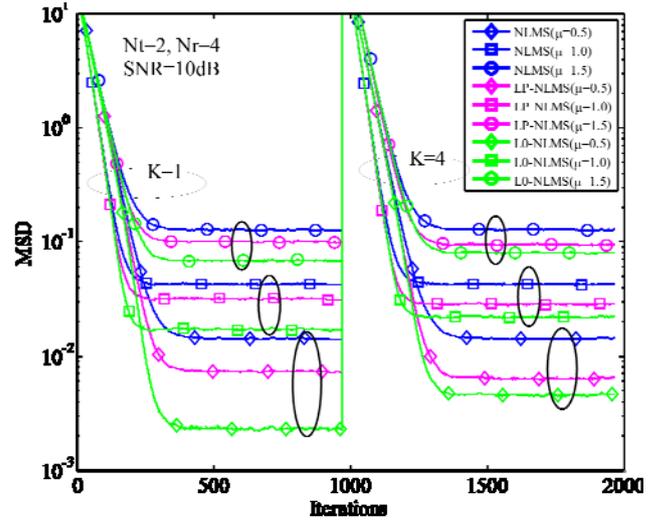

Fig. 10. Performance comparison verses step-size of gradient descend

## V. CONCLUSION

In this paper, we proposed ASCE methods using LP-NLMS and L0-NLMS algorithms, for time-variant MIMO-OFDM systems. First of all, system model was formulated to ensure each MISO channel vector can be estimated independently. Secondly, cost function of the two proposed methods were constructed using sparse penalties, i.e., $L_p$-norm and $L_0$-norm. Later, MIMO channel matrix was estimated using ASCE methods. At last, simulation results were shown that proposed ASCE methods achieved better performance than standard ACE method without scarifying computational complexity.

## ACKNOWLEDGMENT

This work was supported in part by the Japan Society for the Promotion of Science (JSPS) postdoctoral fellowship.

## REFERENCES

[1] D. Raychaudhuri and N. B. Mandayam, "Frontiers of Wireless and Mobile Communications," *Proceedings of the IEEE*, vol. 100, no. 4, pp. 824–840, Apr. 2012.
[2] [F. Adachi, D. Grag, S. Takaoka, and K. Takeda, "Broadband CDMA Techniques," *IEEE Wireless Communications*, vol. 12, no. 2, pp. 8–18, Apr. 2005.
[3] F. Adachi and E. Kudoh, "New direction of broadband wireless technology," *Wireless Communications and Mobile Computing*, vol. 7, no. 8, pp. 969–983, 2007.
[4] J. D. Gibson and R. A. Iltis, "Channel Estimation and Data Detection for MIMO-OFDM Systems," *in IEEE Global Telecommunications Conference (GLOBECOM)*, San Francisco, USA, Dec. 1-5, 2003, pp. 581–585.
[5] M. Biguesh and a. B. Gershman, "Training-based MIMO channel estimation: a study of estimator tradeoffs and optimal training signals," *IEEE Transactions on Signal Processing*, vol. 54, no. 3, pp. 884–893, Mar. 2006.
[6] T. Chang, W. Chiang, and Y. P. Hong, "Training Sequence Design for Discriminatory Channel Estimation in Wireless MIMO Systems," *IEEE*

[7] S. He, J. K. Tugnait, and X. Meng, "On Superimposed Training for MIMO Channel Estimation and Symbol Detection," *IEEE Transactions on Signal Processing*, vol. 55, no. 6, pp. 3007–3021, Jun. 2007.

[8] T. H. Pham, Y. Liang, and A. Nallanathan, "A joint channel estimation and data detection receiver for multiuser MIMO IFDMA systems," *IEEE Transactions on Communications*, vol. 57, no. 6, pp. 1857–1865, Jun. 2009.

[9] Z. J. Wang, Z. Han, and K. J. R. Liu, "A MIMO-OFDM channel estimation approach using time of arrivals," *IEEE Transactions on Wireless Communications*, vol. 4, no. 3, pp. 1207–1213, May 2005.

[10] G. Taubock, F. Hlawatsch, D. Eiwen, and H. Rauhut, "Compressive Estimation of Doubly Selective Channels in Multicarrier Systems: Leakage Effects and Sparsity-Enhancing Processing," *IEEE Journal of Selected Topics in Signal Processing*, vol. 4, no. 2, pp. 255–271, Apr. 2010.

[11] W. U. Bajwa, J. Haupt, A. M. Sayeed, and R. Nowak, "Compressed Channel Sensing: A New Approach to Estimating Sparse Multipath Channels," *Proceedings of the IEEE*, vol. 98, no. 6, pp. 1058–1076, Jun. 2010.

[12] N. Wang, G. Gui, Z. Zhang, and T. Tang, "A Novel Sparse Channel Estimation Method for Multipath MIMO-OFDM Systems," *in IEEE 74th Vehicular Technology Conference (VTC2011-Fall)*, San Francisco, California, USA, Sept. 5-8, 2011, pp. 1–5.

[13] Y. Zhou and A. M. Sayeed, "Experimental Study of MIMO Channel Statistics and Capacity via Virtual Channel Representation," *UW Technical Report, http://dune.ece.wisc.edu/pdfs/zhoumeas.pdf*.

[14] N. Czink, X. Yin, H. OZcelik, M. Herdin, E. Bonek, and B. Fleury, "Cluster Characteristics in a MIMO Indoor Propagation Environment," *IEEE Transactions on Wireless Communications*, vol. 6, no. 4, pp. 1465–1475, Apr. 2007.

[15] L. Vuokko, V.-M. Kolmonen, J. Salo, and P. Vainikainen, "Measurement of Large-Scale Cluster Power Characteristics for Geometric Channel Models," *IEEE Transactions on Antennas and Propagation*, vol. 55, no. 11, pp. 3361–3365, Nov. 2007.

[16] E. J. Candes, J. Romberg, and T. Tao, "Robust Uncertainty Principles : Exact Signal Reconstruction From Highly Incomplete Frequency Information," *IEEE Transctions on Information Theory*, vol. 52, no. 2, pp. 489–509, Feb. 2006.

[17] D. L. Donoho, "Compressed Sensing," *IEEE Transactions on Information Theory*, vol. 52, no. 4, pp. 1289–1306, Apr. 2006.

[18] E. J. Candes, "The restricted isometry property and its implications for compressed sensing," *Comptes Rendus Mathematique*, vol. 1, no. 346, pp. 589–592, May 2008.

[19] A. Amini, M. Unser, and F. Marvasti, "Compressibility of Deterministic and Random Infinite Sequences," *IEEE Transactions on Signal Processing,* vol. 59, no. 99, pp. 5193 - 5201, Nov. 2011.

[20] B. Widrow and D. Stearns, "*Adaptive signal processing,*" New Jersey: Prentice Hall, 1985.

[21] Y. Chen, Y. Gu, and A. O. Hero III, "Sparse LMS for System Identification," *in IEEE International Conference on Acoustics, Speech and Signal Processing (ICASSP)*, Taipei, Taiwan, Apr. 19-24, 2009, pp. 3125–3128.

[22] O. Taheri and S. A. Vorobyov, "Sparse channel estimation with Lp-norm and reweigthted L1-norm penalized least mean square," in *IEEE International Conference on Acoustics Speech and Signal Processing (ICASSP)*, May 22-27, 2011, pp. 2864–2867.

[23] G. Gui, W. Peng, and F. Adachi, "Improved Adaptive Sparse Channel Estimation Based on the Least Mean Square Algorithm," in *IEEE Wireless Communications and Networking Conference (WCNC)*, Shanghai, China, April 4-7, 2013, pp. 1–5.

[24] Y. Gu, J. Jin, and S. Mei, "L0-Norm Constraint LMS Algorithm for Sparse System Identification," *IEEE Signal Processing Letters*, vol. 16, no. 9, pp. 774–777, Sept. 2009.